# The Duopoly Analysis of Graphics Card Market

Nan Miles Xi [1,2]

## Abstract

By analyzing the duopoly market of computer graphics cards, we categorized the effects of enterprise's technological progress into two types, namely, cost reduction and product diversification. Our model proved that technological progress is the most effective means for enterprises in this industry to increase profits. Due to the technology-intensive nature of this industry, monopolistic enterprises face more intense competition compared with traditional manufacturing. Therefore, they have more motivation for technological innovation. Enterprises aiming at maximizing profits have incentives to reduce costs and achieve a higher degree of product differentiation through technological innovation.

## Introduction

This paper discussed the personal computer graphics card market. This market belongs to the field of consumer electronics. After years of competition and mergers and acquisitions, more than 90% of the market share has been divided between NVIDIA and ATI. It is a typical duopoly market structure. The industry's technological advancement and product updates are fast. These enterprises are pursuing profit maximization in the short-term, but in the long-term, they want to increase their market share and squeeze out rivals from the market. The degree of product differentiation is very strong, and the price cut of new products is very fast. Based on the aforementioned industry characteristics, we focused on the impact of product differentiation and cost reduction on the enterprise's decision-making, both of which are rooted in technological progress.

The graphics card industry is typically technology-intensive, and enterprises must always develop new technologies and products to seize the first-mover advantage. Technological progress will bring about two effects, namely cost reduction and product diversification. Mathematically, it can be expressed as

$$\frac{dT}{dt} = -\frac{dC}{dt} + \frac{dD}{dt},$$

where $\frac{dT}{dt}$ is the rate of technological progress; $-\frac{dC}{dt}$ is the rate of production cost decrease induced by technological progress (the negative sign represents the decrease in the cost of producing the same product); $\frac{dD}{dt}$ is the diversification of products induced by technological progress (the positive sign indicates that with technological advancement, new products are introduced continuously). From a practical point of view, oligarchs always have to speed up their own research and development so as to have lower production costs and a richer product line. We will use a model to analyze the reasons for this enterprise behavior and explain that this strategy is the optimal choice for an oligarch whose goal is to maximize profits.

[1] Department of Statistics, University of California, Los Angeles, CA, US
[2] Institute of International Economics, Nankai University, Tianjin, China

## Product differentiation caused by technological progress

### Assumptions about cost

We assume that the enterprise's fixed cost is $C_0$, and the marginal cost is zero. This is because, with the development of semiconductor and integrated circuit technology, the cost of producing a graphics card (circuit board and components) is very low, far below one-tenth of its market price. Therefore, for the convenience of analysis, it can be set to zero without affecting the final result. The main cost of the enterprise is concentrated in the product development process, including the salary of high-tech workers and hardware development equipment. This part of the expenditure constitutes the majority of the enterprise cost, and it is difficult to transfer to other uses once invested, which is fixed Sunk costs.

### Periodic games

We assume that the game process of oligopoly enterprises is indefinite and presents periodic characteristics. Each game cycle is divided into two phases. In the first phase, due to the maturity and diffusion of existing technologies, the products of each enterprise are the same, i.e., homogeneous products. In the second phase, motivated by profit maximization, A enterprise developed new products through technological innovation and then produced product differentiation to obtain greater profits. After B enterprise observed A's behavior, she will also adopt the same strategy. Therefore, product differentiation cannot be maintained forever. New technologies will eventually become mature and public, making the game return to the first stage and form a cycle. The gaming process is shown in Figure 1.

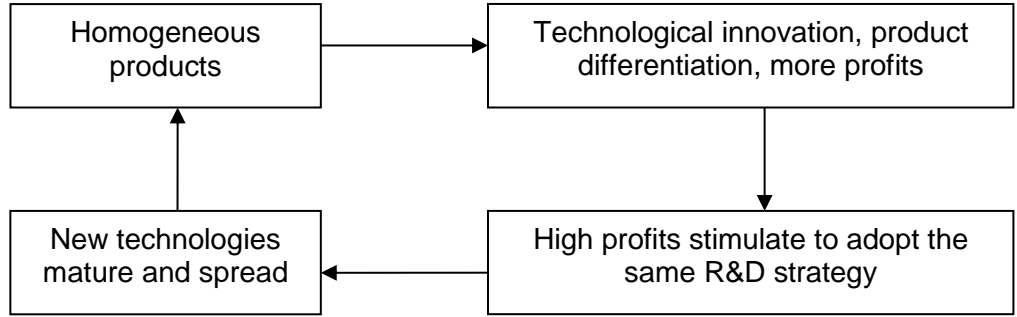

Figure 1. The gaming process of duopoly competition.

**The stage of producing homogeneous products**

Because the products are homogeneous and the marginal cost of production is zero, the game at this stage can be described by the Cournot model. Let $q_A$ and $q_B$ be the output, and $\pi_A$, $\pi_B$ be the profit of enterprise A and B, respectively. A and B both face a downward-sloping market demand curve $p = a - (q_A + q_B)$. Then we have

$$\pi_A = pq_A = aq_A - q_A^2 - q_A q_B$$
$$\pi_B = pq_B = aq_A - q_B^2 - q_A q_B$$

Solving the first-order profit maximization condition

$$\frac{\partial \pi_A}{\partial q_A} = a - 2q_A - q_B = 0$$
$$\frac{\partial \pi_B}{\partial q_B} = a - 2q_B - q_A = 0$$

gives $q_A = q_B = \frac{a}{3}$, $\pi_A = \pi_B = \frac{a^2}{9}$. That is, A and B share the market equally, and their profits are same. This result shows that under the premise of homogenous products, if the market capacity *a* does not change, the profit of the enterprise will remain unchanged, and no enterprise can obtain additional income. The only way to increase profits is to expand market share through advertising or other means. However, if the market capacity increases, then both enterprises benefit. The enterprise that did not advertise may become free riders, which inhibits the enterprise's motivation to increase market capacity. *a* should be regarded as an exogenous variable, decided by changes in national income and consumption habits.

**Product differentiation**

Cournot equilibrium is unstable in the long run. Oligarchs cannot obtain higher profits in Cournot equilibrium and also lose better control over the market (Figure 1). In order to obtain greater profits, enterprises have the motivation to introduce new products through technological progress to achieve differentiation. This is the second stage of the game cycle - the process of differentiation. The Hotelling model is a classic model for describing the product's spatial differentiation. Below we will use the revised Hotelling model to illustrate that product differentiation through technological advancement is the best choice for enterprises under certain assumptions.

We regard the different preferences of consumers as a continuous line. Consumers with different preferences are placed at different points on the line. Consumers with similar preferences are adjacent. Each consumer on the straight line needs one unit of product. If consumers have to consume a product that is not exactly the same as their preferences, they will incur a cost due to dissatisfaction. We assume that this cost is $cx^2$. *x* is the distance between the consumer's favorite product and the supplier closest to their preference. *c* is the cost per unit distance. This assumption indicates that the marginal degree of dissatisfaction is increasing. The model is shown in Figure 2.

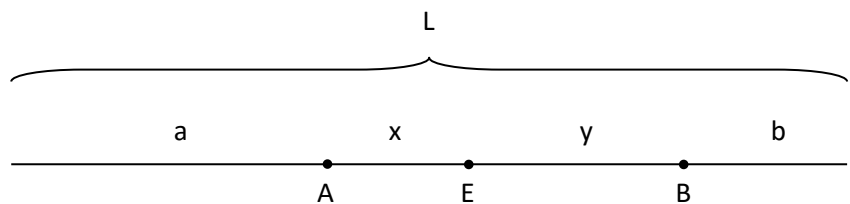

Figure 2. Hotelling model in duopoly competition.

In figure 2, two enterprises A and B are distributed at two points A and B on the straight line L. The distance from A to the left end is a, and the distance from B to the right end is b, which represents the price of enterprise A's and B's products, respectively. For people at point E, purchasing from A and B is the same, that is

$$p_A + cx^2 = p_B + cy^2$$

$$x^2 = \frac{p_B - p_A}{c} + y^2$$

Since $a + x + y + b = L$, then

$$x^2 = \frac{p_B - p_A}{c} + (L - a - b - x)^2$$

$$x = \frac{p_B - p_A}{2c(L-a-b)} + \frac{L^2 + a^2 + b^2 - 2aL - 2bL + 2ab}{2(L-a-b)} \tag{1}$$

Similarly,

$$y = \frac{p_A - p_B}{2c(L-a-b)} + \frac{L^2 + a^2 + b^2 - 2aL - 2bL + 2ab}{2(L-a-b)} \tag{2}$$

Also,

$$\pi_A = p_A(a+x) = p_A[\frac{p_B - p_A}{2c(L-a-b)} + \frac{L^2 - a^2 + b^2 - 2bL}{2(L-a-b)}] \tag{3}$$

$$\pi_B = p_B(b+y) = p_B[\frac{p_A - p_B}{2c(L-a-b)} + \frac{L^2 + a^2 - b^2 - 2aL}{2(L-a-b)}] \tag{4}$$

Solving the first-order profit maximization condition

$$\frac{\partial \pi_A}{\partial p_A} = \frac{p_B - 2p_A}{2c(L-a-b)} + \frac{L^2 - a^2 + b^2 - 2bL}{2(L-a-b)} = 0$$

$$\frac{\partial \pi_B}{\partial p_B} = \frac{p_A - 2p_B}{2c(L-a-b)} + \frac{L^2 + a^2 - b^2 - 2aL}{2(L-a-b)} = 0$$

gives

$$p_A = \frac{c}{3}(3L^2 - a^2 + b^2 - 2aL - 4bL) \tag{5}$$

$$p_B = \frac{c}{3}(3L^2 + a^2 - b^2 - 4aL - 2bL) \tag{6}$$

Plugging (1) and (5) into (3) gives

$$\pi_A = p_A \frac{-a^2 + b^2 - 2aL - 4bL + 3L^2}{6(L-a-b)}$$

Let $E = \dfrac{-a^2 + b^2 - 2aL - 4bL + 3L^2}{6(L-a-b)}$, then

$$\pi_A = p_A E$$

$$\frac{\partial p_A}{\partial a} = \frac{c}{3}(-2a - 2L) \tag{7}$$

$$\frac{\partial E}{\partial a} = \frac{L^2 + a^2 + 2ab - 2bL - 2aL + b^2}{6(L-a-b)} \tag{8}$$

Let $F = L^2 + a^2 + 2ab - 2bL - 2aL + b^2$, since

$$\frac{\partial F}{\partial a} = 2a + 2b - 2L < 0$$
$$F(a=0, b=L) = 0$$

then, monotonicity gives $F < 0$, and (7) and (8) give $\dfrac{\partial p_A}{\partial a} < 0, \dfrac{\partial E}{\partial a} < 0$. Therefore,

$$\frac{\partial \pi_A}{\partial a} = \frac{\partial (p_A E)}{\partial a} = \frac{\partial p_A}{\partial a} E + \frac{\partial E}{\partial a} p_A < 0 \tag{9}$$

Plugging (2) and (6) into (4) gives

$$\pi_B = p_B \frac{a^2 - b^2 - 4aL - 2bL + 3L^2}{6(L-a-b)}$$

The symmetry of Nash equilibrium gives

$$\frac{\partial \pi_B}{\partial b} < 0 \tag{10}$$

Equations (9) and (10) show that the profits of enterprises A and B increase with the decrease of *a* and *b*. When *a=b=0*, A and B reach the two endpoints of the straight line *L* and obtain maximum profits. In other words, when product differentiation reaches the maximum, enterprises gain the greatest profit. This shows that enterprises have the motivation to achieve product differentiation. In technology-intensive industries, only technological innovation can achieve the greatest degree of differentiation because new products would be completely different. Therefore, in reality, we will observe that the development of new products is the most important production strategy of an enterprise.

Now we examine the company's maximum profit. Substitute *a=b=0* into (5) and (6) to get

$$p_A = p_B = cL^2$$

Substitute the above formula into (3) and (4) to get

$$\pi_A = \pi_B = \frac{cL^3}{2}$$

This equation shows that the maximum profit of an enterprise is proportional to the third power of the degree of product differentiation *L*, and is proportional to the degree of dissatisfaction *c* brought by consumers having to consume a product that is not exactly the same as their preferences. Since *c* is an exogenous variable that cannot be changed for an enterprise, then *L* is a decisive factor affecting profits. If an enterprise wants to increase profits, the most powerful means is to increase product differentiation through technological innovation, which is consistent with reality. In summary, for any oligarch, the best strategy is to increase the degree of product differentiation, and the most effective means to achieve this goal is technological progress and innovation in high-tech industries like the graphics card.

**Cost reduction caused by technological progress**

Technological progress can enable manufacturers to use fewer inputs to produce a given output. Therefore, under the premise that the price of input factors remains unchanged, such technological progress moves the total cost curve downward. In the previous section, we assumed that the marginal cost of the firm is zero with a fixed cost $C_0$, making the total cost $C_0$. Suppose an enterprise has a production function $q = A(t)f(k,l)$ with constant returns to scale, where *k* is the capital invested (e.g., equipment for development); *l* is labor (developers); *A(t)* represents technological progress, and is a function of time; *v* is the remuneration of capital and *w* is the remuneration of the developers. Note that the cost function is a homogeneous function of output. Then the total cost in the initial period can be written as

$$C_0 = C_0(v, w, q) = qC_0(v, w, 1)$$

Because the input to produce a unit of product in *t=0* will produce a unit of product in *t*, we have

$$C_t(v, w, A(t)) = A(t)C_t(v, w, 1) = C_0(v, w, 1)$$

We can obtain the total cost function in period *t*

$$C_t(v, w, q) = qC_t(v, w, 1) = \frac{qC_0(v, w, 1)}{A(t)} = \frac{C_0(v, w, q)}{A(t)}$$

Therefore,

$$C_t(v, w, q) < C_0(v, w, q)$$

It can be seen that even if product differentiation is not carried out, only the reduction of production costs caused by technological progress can increase the profit of the enterprise (under the original price). In the high-tech industry, technological advancement not only brings improvements in production efficiency (thereby reducing costs and increasing profits) but also promotes the formation of new products (which in turn increases differentiation and profits). Therefore, the common incentives of these two aspects will make enterprises prefer technological innovation more than traditional manufacturing, and the degree of competition will become more intense. So far, we can sum up the payment matrix of the game between two enterprises in figure 3.

The strategy of NVIDIA

| The strategy of ATI | | R&D | No R&D |
|---|---|---|---|
| | R&D | 50, 50 | 200, 0 |
| | No R&D | 0, 200 | 100, 100 |

Figure 3 The payment matrix of duopoly competition

## Conclusion

As shown in Figure 3, (R&D, R&D) is the Nash equilibrium solution, and the result of the game between the two enterprises is similar to the "Prisoner's Dilemma." But it is not a "dilemma" for consumers because consumers have obtained a wealth of products and relatively low prices from fierce competition.